\documentclass{aa}  
\usepackage{graphicx}
\usepackage{txfonts}

\usepackage{ulem}

\def\farcs{\hbox{$.\!\!^{\prime\prime}$}}  
\def\asec{\ifmmode ^{\prime\prime}\else$^{\prime\prime}$\fi}

\begin{document}

    \title{X-ray Flashes or soft Gamma-ray Bursts?}

    \subtitle{The case of the likely distant XRF 040912}

   \author{G. Stratta
          \inst{1}
        \and S. Basa\inst{2}
        \and N. Butler\inst{3}
       \and J.L. Atteia\inst{1}
        \and B. Gendre\inst{4}
        \and A. P\'elangeon\inst{1}
        \and F. Malacrino\inst{1}
        \and Y. Mellier\inst{5}
        \and D. A. Kann\inst{6}
        \and S. Klose\inst{6}
       \and A. Zeh\inst{6}
        \and N. Masetti\inst{7}
        \and E. Palazzi\inst{7}
	\and J. Gorosabel\inst{8}
	\and A.J. Castro-Tirado\inst{8}
	\and A. de Ugarte Postigo\inst{8}
	\and M. Jelinek\inst{8}
	\and J. Cepa\inst{9}
	\and H. Casta\~neda\inst{9}
	\and D. Mart\'{\i}nez-Delgado\inst{9}
        \and M. Bo\"er\inst{10}
        \and J. Braga\inst{11}
        \and G. Crew\inst{12}
        \and T.~Q.~Donaghy\inst{13}
        \and J-P. Dezalay\inst{14}
        \and J. Doty\inst{12}
        \and E.~E.~Fenimore\inst{15}
        \and M.~Galassi\inst{15}
        \and C.~Graziani\inst{13}
        \and J.G. Jernigan\inst{16}
        \and N. Kawai\inst{17,18}
        \and D. Q. Lamb\inst{13}
        \and A. Levine\inst{12}
        \and J. Manchanda\inst{19}
        \and F. Martel\inst{12}
        \and M.~Matsuoka\inst{20}
        \and Y. Nakagawa\inst{22}
        \and J-F. Olive\inst{14}
        \and G. Pizzichini\inst{7}
        \and G. Prigozhin\inst{12}
        \and G. Ricker\inst{12}
        \and T. Sakamoto\inst{25}
        \and Y. Shirasaki\inst{18,21}
        \and S. Sugita\inst{22}
        \and M. Suzuki\inst{18}
        \and K. Takagishi\inst{23}
        \and T. Tamagawa\inst{18}
        \and R. Vanderspek\inst{12}
        \and J. Villasenor\inst{12}
        \and S.E. Woosley\inst{24}
        \and M. Yamauchi\inst{23}
        \and A. Yoshida\inst{18,22}
          }

   \offprints{G. Stratta}

\institute{
Laboratoire d'Astrophysique de Toulouse, Observatoire Midi-Pyr\'en\'ees, 14 Ave. E. Belin, 31400 Toulouse, France.
\and Laboratoire d'Astrophysique de Marseille, Traverse  du Siphon-Les trois Lucs, 13012, Marseille, France
\and Space Sciences Laboratory, University of California at Berkeley, 445 Campbell Hall, Berkeley, CA 94720, USA
\and INAF/IASF Roma, via fosso del Cavaliere 100, 00133 Roma, Italy
\and Institut d'Astrophysique de Paris, 98 bis Boulevard Arago, 75014, Paris, France
\and Th\"uringer Landessternwarte Tautenburg, Sternwarte 5, D - 07778 Tautenburg, Germany
\and INAF/IASF Bologna, via Gobetti 101, 40129 Bologna, Italy
\and Instituto de Astrof\'isica de Andaluc\'ia (IAA-CSIC), Apartado de Correos 3004, 18080 Granada, Spain 
\and Instituto de Astrof\'isica de Canarias (IAC), C. Via L\'actea s/n, La Laguna, E-38200, Tenerife, Spain
\and Observatoire Haute Provence (CNRS/OAMP), Saint Michel l'Observatoire, France
\and Instituto Nacional de Pesquisas Espaciais, Avenida Dos Astronautas 1758, S\~ao Jos\'e dos Campos 12227-010, Brazil.
\and Center for Space Research, MIT, 70 Vassar Street, Cambridge, MA 02139
\and Department of Astronomy and Astrophysics, University of Chicago, 5640 South Ellis Avenue, Chicago, IL 60637.
\and Centre d'Etude Spatiale des Rayonnements, Observatoire Midi-Pyr\'{e}n\'{e}es, 9 Avenue de Colonel Roche, 31028, Toulouse, France
\and Los Alamos National Laboratory, P.O. Box 1663, Los Alamos, NM, 87545.
\and University of California at Berkeley, Space Sciences Laboratory, Berkeley, CA, 94720-7450.
\and Department of Physics, Tokyo Institute of Technology, 2-12-1 Ookayama, Meguro-ku, Tokyo 152-8551, Japan.
\and RIKEN (Institute of Physical and Chemical Research), 2-1 Hirosawa, Wako, Saitama 351-0198, Japan.
\and Department of Astronomy and Astrophysics, Tata Institute of Fundamental Research, Homi Bhabha Road,Mumbai, 400 005, India.
\and Tsukuba Space Center, National Space Development Agency of Japan, Tsukuba, Ibaraki, 305-8505, Japan.
\and National Astronomical Observatory, 2-21-1, Osawa, Mitaka, Tokyo, 181-8588
\and Department of Physics, Aoyama Gakuin University, Chitosedai 6-16-1 Setagaya-ku, Tokyo 157-8572, Japan.
\and Faculty of engineering, Miyazaki University, Gakuen Kibanadai Nishi, Miyazaki 889-2192, Japan.
\and Department of Astronomy and Astrophysics, University of California at Santa Cruz, 477 Clark Kerr Hall, Santa Cruz, CA95064.
\and NASA Goddard Space Flight Center, Greenbelt, MD, 20771.
             }

   \date{Received; accepted}

 
  \abstract
   {The origin of X-ray Flashes (XRFs) is still a mystery and several models have been proposed.
   To disentangle among these models, an important observational tool is the measure of the XRF distance scale, so far available only for a few of them.}
   {In this work, we present a multi-wavelength study of XRF 040912, aimed at measuring its distance scale and the intrinsic burst properties. }
   {We performed a detailed spectral and temporal analysis of both the prompt and the afterglow emission and we estimated the distance scale of the likely host galaxy. We then used the currently available sample of XRFs with known distance to discuss the connection between XRFs and classical Gamma-ray Bursts (GRBs).}	
   {We found that the prompt emission properties unambiguously identify this burst as an XRF, with an observed peak energy of $E_p=17\pm13$ keV and a burst fluence ratio $S_{2-30keV}/S_{30-400keV}>1$. A non-fading optical source with $R\sim24$ mag and with an apparently extended morphology is spatially consistent with the X-ray afterglow, likely the host galaxy. XRF 040912 is a very dark burst since no afterglow optical counterpart is detected down to $R>25$ mag (3$\sigma$ limiting magnitude) at 13.6 hours after the burst.  The host galaxy spectrum detected from 3800 \AA~to 10000 \AA, shows a single emission line at 9552 \AA. The lack of any other strong emission lines blue-ward of the detected one and the absence of the Ly$\alpha$ cut-off down to 3800~\AA~are consistent with the hypothesis of the [OII] line at redshift $z=1.563\pm0.001$. 
The intrinsic spectral properties rank this XRF among the soft GRBs in the $E_{peak}-E_{iso}$ diagram. Similar results were obtained for most XRFs at known redshift. Only XRF 060218 and XRF 020903 represent a good example of instrinsic XRF ({\it i}-XRF) and are possibly associated with a different progenitor population.  This scenario may calls for a new definition of XRFs. 
}
   {}

   \keywords{gamma-ray : bursts 
              }

   \maketitle

%

\section{Introduction}

X-ray Flashes (XRFs; Heise et al. \cite{Heise2001}; Kippen et al. \cite{Kippen2001}) are extra-galactic transient X-ray sources with spatial distribution, spectral and temporal characteristics similar to long duration Gamma-Ray Bursts (GRBs).
The remarkable property that distinguishes XRFs from GRBs is that their $\nu F_{\nu}$ prompt emission spectrum 
peaks at energies which are observed to be typically one order of magnitude lower than the observed peak energies of GRBs.
XRFs are empirically defined by a greater fluence (time integrated flux) in the X-ray band (2-30 keV) than in the gamma-ray band (30-400 keV). Along with the intermediate class of bursts called X-ray Rich bursts (XRRs), XRFs and GRBs form a continuum in the observed peak energy versus the 2-400 keV fluence  plane, defining the well known hardness-intensity relation for which soft bursts are also faint, in the observer reference frame (Barraud et al. \cite{Barraud2003}, Sakamoto et al. \cite{Sakamoto2005}).
The XRF optical and X-ray afterglow flux distribution at about 0.5 days from the burst trigger is comparable to the one observed for GRB afterglows (D'Alessio et al. \cite{Dalessio2005}). Despite the lack of any clear evidence of achromatic temporal breaks so far, XRF afterglow temporal decay indices are on average consistent with those commonly observed for GRB and XRR afterglows  before or after the temporal break (D'Alessio et al. \cite{Dalessio2005}).

As soon as they were discovered, XRFs were thought to be high redshift ($z\ga5$) GRBs, so that their softness and faintness were an effect of the cosmological distance (Heise et al. 2003). However, the absence of any evidence of time dilation in the temporal properties of XRFs with respect to nearby GRBs motivated other explanations. The origin of XRFs has been explored in the context of the fireball model, where bursts fainter and softer than GRBs can be produced by: {\it i)} high baryon-loaded fireballs, assuming that both the prompt emission and the afterglow are produced by external shocks (``dirty fireball", e.g. Dermer et al. \cite{Dermer1999}); {\it ii)} low-efficiency internal shocks due to low-velocity contrast between the high velocity colliding shells (e.g. Mochkovitch et al. \cite{Mochkovitch2004}, Barraud et al. \cite{Barraud2005}).
Alternatively, it has been proposed that XRFs and GRBs are indeed the same phenomenon. In this case, their diversity is associated to the aperture of the jet opening angle (Jet Variable Opening Angle model) or to the observer viewing angle from the jet axis, for a structured energy density distribution within the jet (Structured Jet model; see Lamb et al. \cite{Lamb2005} for a review). In the so-called ``off-axis" GRB model, the observer line-of-sight is outside the jet cone. In this case, the prompt emission is detectable only when the relativistic beaming angle is large enough to enter the line-of-sight (e.g. Yamazaki et al. \cite{Yamazaki2002}). Alternatively, an XRF is observed in a sideway vision at the moment when the jet is breaking out from a hot cocoon surrounding the GRB (e.g. M\'esz\'aros et al. \cite{Meszaros2002}).

An important observational tool to disentangle the proposed models is the measure of the distance scale of XRFs and therefore the intrinsic properties of the burst.
Among 35 XRFs observed so far (from BeppoSAX and HETE-2 archives, see D'Alessio et al. \cite{Dalessio2005} and Sakamoto et al. \cite{Sakamoto2005}, and from J. Greiner's web page\footnote{\rm{http://www.mpe.mpg.de/$\sim$jcg/grbgen.html}} from September 2003 to August 2006), the ones at known redshift are \object{XRF 020903} at 
$z=0.2506\pm0.0003$ (Bersier et al. \cite{Bersier2006}), \object{XRF~030429} at $z=2.658\pm0.004$ (Jakobsson et al. \cite{Jakobsson2004}), \object{XRF~030528} at $z=0.782\pm0.001$ (Rau et al. \cite{Rau2005}), \object{XRF~040701} at $z=0.2146$ (Kelson et al. \cite{Kelson2004}), \object{XRF~050416A} at $z=0.6528\pm0.0002$ (Soderberg et al. \cite{Soderberg2006}), \object{XRF~050824} at $z=0.83$ (Crew et al. \cite{Crew2005}, Fynbo et al. \cite{Fynbo2005}) and \object{XRF 060218} at $z=0.03342$ (Pian et al. \cite{Pian2006}).  Recently, \object{GRB~981226}, detected with the BeppoSAX satellite, has been reanalyzed by D'Alessio et al. (\cite{Dalessio2005}) and classified as an XRF, at $z=1.11\pm0.06$ (Christensen et al. \cite{Christensen2005}). As already pointed out by Rau et al. (\cite{Rau2005}), some XRFs (e.g. \object{XRF~030528}, \object{XRF~030429}) would have been detected as XRR bursts or GRBs if observed in their rest frame, possibly suggesting that some XRFs do not require a different emission model than for GRBs.

In this paper we present and discuss the results obtained from a multi-wavelength analysis of \object{XRF 040912} aimed at measuring its distance scale and the intrinsic burst properties. In \S 2 we provide a brief review of the follow-up campaign of \object{XRF 040912}. In \S3 we present the data reduced in this work and the performed analysis. In \S4 we present the results obtained from our analysis while in \S5 we discuss the dark nature of this burst and we compare the intrinsic properties 
of XRF 040912 with those observed for other XRFs and GRBs. In \S6 we discuss the implications of our observations for the nature of X-ray flashes.


\section{Observations}

\object{XRF 040912} was a soft, long burst discovered with the Wide-Field X-Ray Monitor (WXM) and the French Gamma Telescope (FREGATE) on board the {\it High Energy Transient Explorer-2} (HETE-2) on 2004, September 12.592 UT (Butler et al. \cite{Butler2004a}).
The $7'$ WXM error-box was centered at R.A.= $23^h 56^m 53.^s52$ and Dec.$=-01^{\circ} 00^{'} 03\farcs6$ (J2000.0). Optical follow-up started 1.8 hours after the burst (Ogura et al. 2004) with several ground based telescopes but, despite very deep observations, no evidence of a transient source was found. The Chandra Observatory targeted the WXM error-box in two epochs, from 3.32 to 3.57 days and from 8.86 to 9.12 days after the burst (Butler et al. \cite{Butler2004b}). Among 22 X-ray sources found, only one  (CXOU J235642.9--005520, hereafter S1) showed a significantly fading behavior. 
Two R-band observations of the HETE-2 WXM error region were performed in two epochs with the $27'\times27'$ FOV IMACS camera on the Magellan 6.5m Baade Telescope at Las Campanas Observatory (Butler et al. \cite{Butler2004c}), with $3\times180$ s exposures at 13.57 hours after the burst and $2\times300$ s exposures at 38.65 hours after the burst (mean observing time). 
From these observations a persistent optical source spatially consistent with the fading X-ray source S1 was found, while an apparently fading optical source was found positionally consistent with the Chandra source CXOU J235656.4--005839 (hereafter S2) that was non-fading in X-rays (Butler et al. \cite{Butler2004c}). 
Further optical images have been acquired (Gorosabel et al. \cite{Gorosabel2004}) starting 7.25 hr after the event with the 4.2m William Herschel Telescope (WHT) equipped with the Prime Focus Camera (PFC) at La Palma ($16'\times 16$' FOV) and 0.5 days after the burst with the Sierra Nevada Observatory (OSN) 1.5m Telescope. Internal comparison performed between a sequence of V-band images, within 7.7 hours and 3.5 days after the burst, did not reveal any variability  in the entire HETE-2 error-box (Gorosabel et al. \cite{Gorosabel2004}). 

From the Tautenburg 1.34m Schmidt telescope (TLS) $BVR_CI_C$ observations were obtained between 5 and 11 hours after the burst. In addition, $R_C$ and $I_C$ observations and then $I_C$ observations were obtained one day later and three days after the burst, respectively (Klose et al. 2004), with no evidence of transient source. 
On 2004 Sept. 23.24 UT (10.6 days after the burst), Very Large Array follow-up observations at 8.46 GHz did not reveal any radio counterpart of the X-ray afterglow candidate down to a 3$\sigma$ limit of 120 $\mu$Jy (Soderberg et al. \cite{Soderberg2004}).  On October 5.42 UT and on November 18.35 UT (22.8 and 66.76 days after the burst) we observed the X-ray afterglow candidate with the wide-field imager MegaCam (36 2048 $\times$ 4612 pixel CCDs, 1 square degree FOV), mounted on the 3.6m Canadian-French-Hawaii Telescope (CFHT), with $i'$-band $2\times860$ s exposures, in order to look for any SN emission (Stratta et al. \cite{Stratta2004a}). We obtained further I-band observations with the 8.2m Very Large Telescope (VLT) equipped with Focal Reducer Spectrometer FORS2 on the 30th of September and on the 4th of October 2005. Imaging and spectroscopy of the likely host galaxy were performed under the Director's Discretionary Time Proposal 275.A-5041.

\section{Data reduction and analysis}

Spectra and light curves of the prompt emission were extracted from the HETE-2/WXM (2-25 keV) and HETE-2/FREGATE (6-400 keV) data. The spectral analysis was performed with a joint WXM and FREGATE data fit. The spectra have been integrated over 146.8 s, that is from $T_0-12.7$ s to $T_0+134.1$ s, where $T_0$ is the HETE-2 trigger time. The joint spectral fits WXM/FREGATE have been performed within the 5-200 keV energy range, using the XSPEC v.11.2.0 software package.

Spectral analysis was performed for each Chandra Target of Opportunity observations (ToO), with integration time of 18.2 ks and 19.5 ks  respectively, for the two sources that showed a fading behavior (S1 and S2, see \S2). We reprocessed the data using the latest packages available (CIAO v 3.3.0.1 and caldb v 3.2.1) and the standard procedures described on the CIAO webpage. 

Astrometry and photometry of the Magellan, TLS, WHT, OSN and VLT/FORS2 images have been performed following standard procedures. Magnitudes have been calibrated against the USNO stars of Henden et al. (2004). Photometry was performed using the {\rm SExtractor} software (Bertin \& Arnouts \cite{Bertin1996}). In order to check for flux variability, the two epochs Magellan images have been subtracted with the dedicated package ISIS (Alard \cite{Alard2000}).  For the CFHT/MEGACAM images, flat-field and bias correction were performed with the {\rm Elixir} pre-processing tool at the CFHT. Quality assessment of individual and stacked images, astrometric calibration, image re-centering and the full stacked process, were performed at the Terapix astronomical data reduction center. Unfortunately, the poor quality of the second CFHT/MEGACAM epoch observations (see \S2) prevented us to check for the presence of any supernova component through flux variability.  VLT/FORS2 long slit spectroscopy has been performed with $4\times1100$ s exposures using the G150I grism, covering the spectral range of 3500$\div$10000 \AA. Spectra have been reduced by using an automatic pipeline especially developed in the framework of The SuperNova Legacy Survey (Basa et al. in preparation).

\section{Results}

   \begin{table}
      \caption[]{Burst duration at different energy ranges of \object{XRF 040912}}
         \label{t:tlc}
     $$ 
         \begin{array}{p{0.5\linewidth}lll}
            \hline
            \noalign{\smallskip}
            Instrument      & \Delta E[keV] & T_{90}[s] & T_{50}[s] \\
            \noalign{\smallskip}
            \hline
            \noalign{\smallskip}
            HETE-2/WXM       & 2-25   &  143\pm10 &  70.0\pm9.5 \\
	    HETE-2/FREGATE   & 6-40   &  127\pm13 & 80\pm5      \\
                 	     & 30-400 &   104\pm26 & 53\pm19      \\

            \noalign{\smallskip}
            \hline
         \end{array}
     $$ 
   \end{table}
%

%
\begin{table*}
\caption{Results from WXM (5-25 keV) and FREGATE (7-200 keV) joint spectral analysis. }             
\label{t:tspec}      
\centering          
\begin{tabular}{lccccccc}   
\hline\hline       
Model   & $\Gamma_1$  & $\Gamma_2$ &   $E_0$ (keV)  &  $E_{peak}$ (keV) & Norm. WXM     &   Norm. FREGATE
      &      $\chi^2_{\nu}/dof$\\
\hline                    
PL$^a$      &  -        &  $-2.15^{+0.22}_{-0.19}$  &   -        &     -   &   $1.56^{+0.96}_{-0.57}$   &  $2.37^{+1.79}_{-1.00}$  & 1.109/104\\

CPL 1$^b$   & $-1.25^{+0.83}_{-0.58}$      &      -         &  $26.98^{+52.85}_{-15.58}$ & -  &   $0.34^{+0.62}_{-0.25}$ & $0.39^{+0.64}_{-0.27}$ & 1.035/103\\

CPL 2$^c$   &  $-1.25^{+0.82}_{-0.58}$   &  -  &  - &  $20.23^{+9.54}_{-5.39}$  & $(1.14^{+1.70}_{-0.53})10^{-2}$ & $(1.53^{+2.09}_{-0.61})10^{-2}$ & 1.035/103\\

Band    &  $-1.18^{+1.23}_{-1.16}$ &   $-3.0$(fixed)    &    $22.61 ^{+73.53}_{-15.02}$   &   - &  $(0.14^{+5.81}_{-0.12})10^{-2}$ &  $(0.18^{+7.38}_{-0.10})10^{-2}$ &1.104/103\\

\hline                  
\end{tabular}
\begin{list}{}{}
\item[$^{a}$] PL = Power-law model;
\item[$^{b}$] CPL 1 = Cutoff power-law model with photon index, $E_0$ and normalization at 1 keV;
\item[$^{c}$] CPL 2 = Cutoff power-law model with photon index, $E_{peak}=(2+\alpha)E_0$ and normalization at 15 keV.
\end{list}

\end{table*}
%

%
   \begin{figure}
   \centering
   \includegraphics[width=9cm]{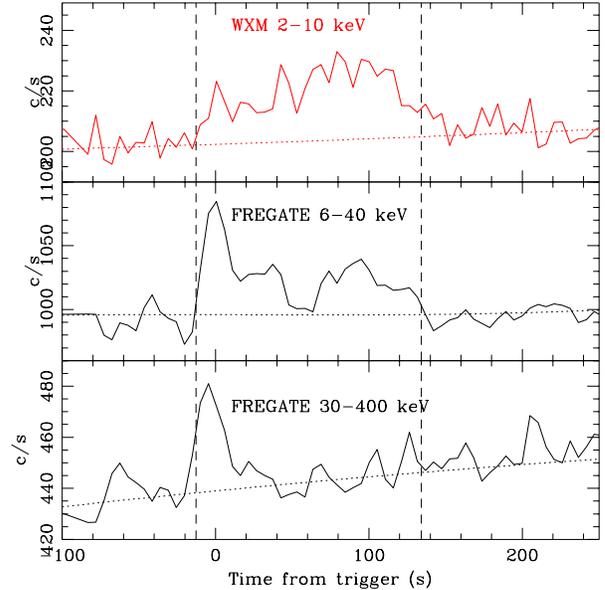}
      \caption{The count rate evolution over the time after the trigger of the prompt emission (temporal resolution of 4 s per bin). The two vertical dashed lines mark the temporal interval in which we extracted the energy spectrum of the burst from the WXM and FREGATE data. The dotted line is a fit of the background count rate evolution with time.}
         \label{f:lc}
   \end{figure}

\subsection{Prompt emission}

The count rate light curve of the prompt emission of \object{XRF 040912} shows a long and soft burst (Fig. \ref{f:lc}). 
We computed the burst durations, represented with the $T_{90}$ and $T_{50}$ parameters, in different energy bands (Tab. \ref{t:tlc}, quoted errors are at 1$\sigma$ level). The 5-200 keV burst spectrum is adequately fit by a simple power-law model or by a cut-off power-law model (Tab. \ref{t:tspec}). The Band model (Band et al. \cite{Band1993}) does not allow to well constrain the parameters. We found a minimum $\chi^2$ by fixing the high energy photon index $\Gamma_2$ to $-3.0$ (Tab. \ref{t:tspec}). Fixing the low energy photon index $\Gamma_1$ to the best fit value obtained from the cut-off power-law model and $\Gamma_2$ to the typical value observed for GRBs of $-2.3$ (Sakamoto et al. 2005), we obtain a peak energy of $E_{peak}=17\pm13$ keV and X-ray and gamma-ray fluences of $S_{2-30 keV}=9.5\times 10^{-7}$ erg cm$^{-2}$ and $S_{30-400 keV}=7.4\times10^{-7}$ erg cm$^{-2}$. The low value of the derived peak energy and the fluence ratio $S_{2-30keV}/S_{30-400keV}>1$  unambiguously identify this burst as an X-ray flash.

\subsection{Afterglow identification}

%
   \begin{figure}
   \centering
   \includegraphics[width=9cm]{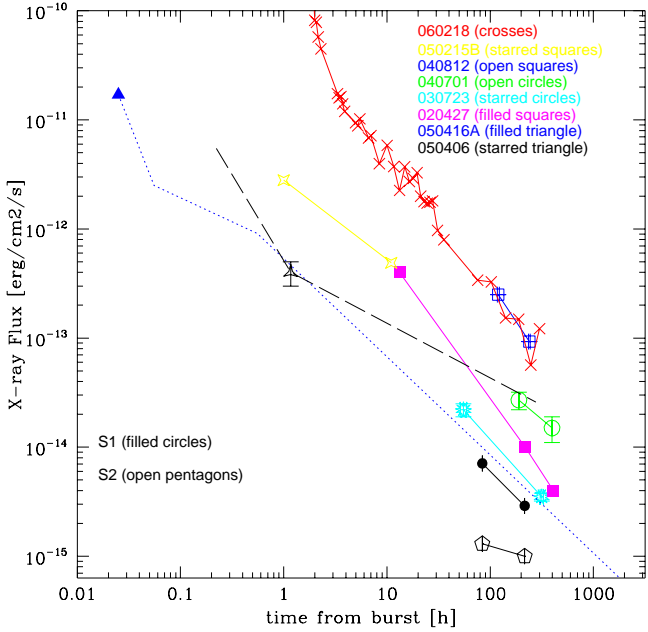}
      \caption{X-ray afterglow light curves of a sample of XRFs for which we could trace the decaying behavior (from GCNs), and the two afterglow candidates of \object{XRF 040912} S1 and S2 (see \S2). For \object{XRF 050416A}, we plot the first measurement and we trace the decay behavior derived by Mangano et al. (\cite{Mangano2006}). For \object{XRF 050406} we plot the flux at the break and the best fit broken power law derived by Romano et al. (\cite{Romano2006}). Other data are from: Campana et al. (\cite{Campana2006}) for 
\object{XRF 060218}, Levan et al. (\cite{Levan2006}) for \object{XRF 050215B}, Campana et al. (\cite{Campana2004}) for \object{XRF 040812}, Fox et al. (\cite{Fox2004}) for \object{XRF 040701}, Butler et al. (\cite{Butler2004d}) for \object{XRF 030723}, Fox et al. (\cite{Fox2002}) for \object{XRF 020427}. While S1 (filled circles) shows a typical temporal behavior, S2 (open pentagons) shows an anomalous decay. }
         \label{f:Xlc}
   \end{figure}
%

From refined photometric analysis of the Magellan Telescope images, we find that the fading behavior of the optical source associated with S2 (Butler et al. 2004c, see \S2) is not real, being an artifact due to bad calibration that brought to a flux overestimation of this particular source during the first epoch. 
We further confirm this result also from the TLS R-band observations, from which we detect the source in three epochs (0.270, 0.431 and 1.353 days after the burst event) with flux constant with time. From the Magellan image subtraction, we find no evidence of any residual for both afterglow candidates S1 and S2 down to $R>25.3$ mag ($3\sigma$ limiting magnitude). 
The afterglow identification is therefore addressed to the X-rays observations. 

The Chandra source S1 (see \S2) is the only one that showed a statistically significant fading behavior in the X-rays. The extracted 0.2-10.0 keV spectra have been fitted with a simple power-law model plus an absorption  component. During the first ToO, the spectral properties of this source show an equivalent neutral hydrogen column density $N_H$ consistent with the Galactic value of $3.7\times10^{20}$cm$^{-2}$ (Dickey \& Lockman \cite{Dickey1990}), an energy spectral index of $\alpha=1.4^{+1.0}_{-0.9}$ and a 2-10 keV flux of $(7.1\pm1.5)\times10^{-15}$ erg cm$^{-2}$ s$^{-1}$. These values are consistent with those estimated for other X-ray afterglows observed at similar epochs (e.g. De Pasquale et al. 2006). During the second ToO the statistics is too poor to constrain the spectral parameters. Assuming the best fit model found in the first ToO and leaving the normalization free to vary, the 2-10 keV flux is $(2.9\pm1.0)\times10^{-15}$ erg cm$^{-2}$ s$^{-1}$ implies a decay index of $\delta_X=-1.0\pm0.6$ consistent with the previous findings of Ford et al. (\cite{Ford2004}; quoted errors are at 90$\%$ confidence level).

For the source S2, the spectral model is poorly constrained due to the faintness of this source.
Assuming the same spectral parameters derived for S1, we found a flux of $(1.3 \pm 0.5) \times 10^{-15}$ erg s$^{-1}$ cm$^{-2}$ and $(1.0 \pm 0.6) \times 10^{-15}$ erg s$^{-1}$ cm$^{-2}$ in the first and second ToO respectively. The corresponding decay index is then $\delta_X = -0.3^{+1.0}_{-1.3}$ ($90\%$ confidence level), confirming previous analysis (Ford et al. \cite{Ford2004}) that classified this source as consistent with being constant in flux with time.

We conclude that, despite the unknown probability of finding a transient source within $7'$ error region with the observed decay index, since S1 is the only Chandra source among 22 detected in the HETE-2 error-box that shows a statistically significant fading behavior in X-rays and both its spectral and temporal properties are consistent with the typical X-ray afterglows observed at similar epochs (few days after the burst), this source provides convincing evidence to be the X-ray afterglow of \object{XRF 040912}. 

We find no fading optical counterpart for this source from a set of several optical images taken at different telescopes and at different epochs (see Tab.\ref{t:ul}), ranking this burst among the most darkest XRF ever observed (Fig. \ref{f:Raft}).

   \begin{table}
      \caption[]{Optical detections and upper limits at the position of the X-ray afterglow candidate. Magnitudes have been corrected for the Galactic extinction (E(B-V)=0.028, Schlegel et al. 1998).}
         \label{t:ul}
     $$ 
         \begin{array}{p{0.2\linewidth}lll}

            \hline
            \noalign{\smallskip}
            Telescope      & \Delta T^a  &  Exp. & mag  \\
			  &    days	& s	   &  	   \\
            \noalign{\smallskip}
            \hline
            \noalign{\smallskip}
TLS              &  0.270  & 1800  & R_C> 22.6 \\
TLS              &  0.289  & 1800  & I_C> 21.0 \\
TLS	  	 &  0.309  &  1800 & B > 22.7\\
WHT		 &  0.312  &  1455 & B > 25.0 \\
WHT		 & 0.325   &  900  & V > 24.4 \\
TLS		 &  0.328  & 1800  & V > 22.9 \\
TLS              &  0.431  & 3360  & R_C> 22.9 \\
OSN		 &  0.553  & 540   & I >  21.0 \\
Magellan$^b$	 &  0.565  & 540   & R=23.89\pm0.10\\
WHT		 &  0.599  & 900   & V > 24.7\\
WHT		 &  1.294  & 1200  & V > 24.7\\
TLS              &  1.353  & 4320  & R_C> 23.3 \\
TLS              &  1.367  & 3600  & I_C> 21.2 \\
Magellan$^b$	 &  1.610  & 600   & R=24.07\pm0.10\\
TLS              &  3.308  & 2700  & I_C> 21.0\\
WHT      	 & 3.489   & 1500  & I > 22.1 \\
CFHT          	 &  22.8   & 1720  & i'=24.02\pm0.16 \\
CFHT          	&  66.7    & 1720  & i'>22.8 \\
VLT             & 385 	   & 690    & I=23.96\pm0.10 \\
            \noalign{\smallskip}
            \hline
         \end{array}
     $$ 
\begin{list}{}{}
\item[$^a$] $\Delta T$ is the time from the burst event.
\item[$^b$] From image subtraction, we find no residual down to $R>25.3$ ($3\sigma$ upper limit).
\end{list}
   \end{table}

\subsection{The likely host galaxy}

\begin{figure}
   \centering
   \includegraphics[width=9cm]{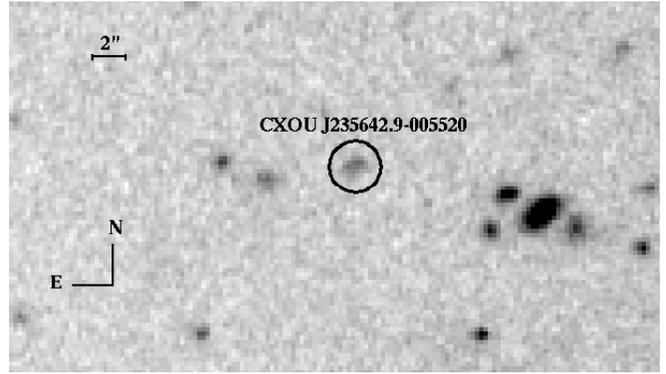}
      \caption{The I-band VLT/FORS2 image of the likely host galaxy of \object{XRF 040912} one year after the burst. The black circle is the Chandra error region of the X-ray afterglow candidate.
}
         \label{f:hostima}
   \end{figure}

The non-fading nature of the optical source positionally consistent with the X-ray afterglow candidate and its apparently extended morphology, strongly suggest that this source is the host galaxy of \object{XRF 040912} (Fig. \ref{f:hostima}).
We computed the chance probability to find a galaxy with R$\le24.0$ mag within the $1''$ radius error circle. Following Piro et al. (2002), we found P=$1.8\times10^{-2}$. This value makes the chance of association unlikely, though not negligible.

\begin{figure*}
   \centering
   \includegraphics[width=12cm]{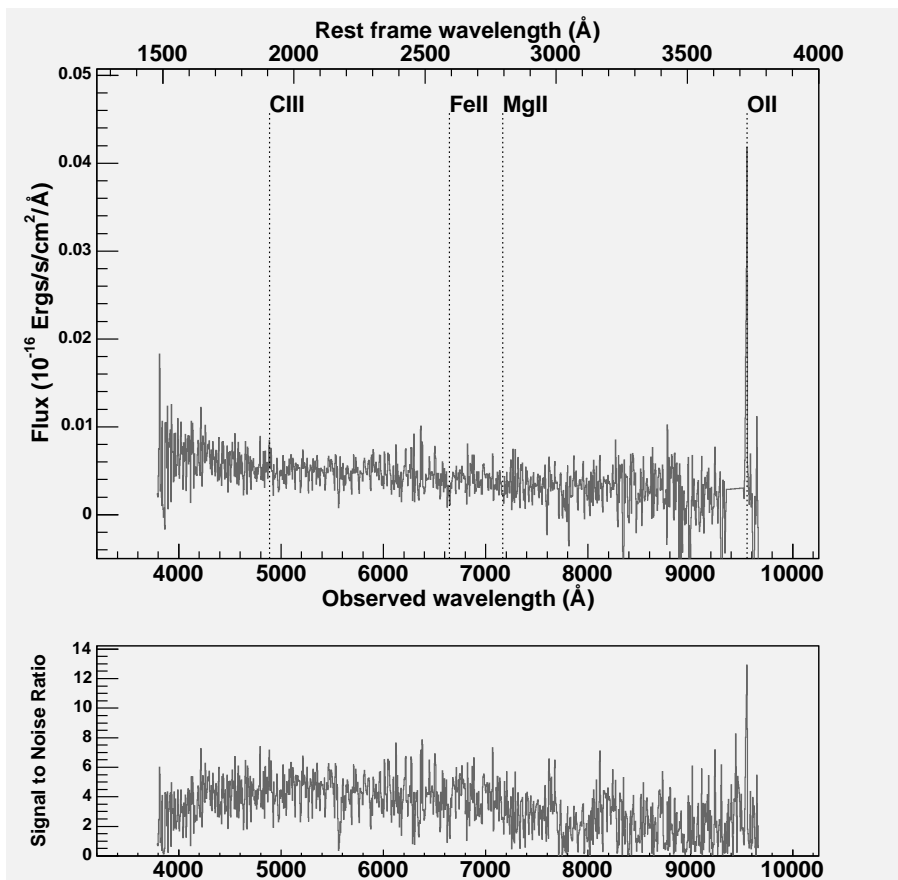}
   \caption{The likely host galaxy VLT/FORS2 1-D spectrum (upper panel) and the associated error spectrum (lower panel). The bad sky subtraction at 9400-9500 \AA~has been hidden. The line identifications imply a redshift $z=1.563\pm0.001$.
}
              \label{f:hostspec}%
    \end{figure*}

The VLT/FORS2 extracted spectrum of this source is well detected between 3800 \AA~and 10000 \AA~, implying a robust redshift limit of $z<2.13$ from the absence of the Ly$\alpha$ absorption due to the neutral hydrogen present along the line-of-sight. The most statistically significant feature is an emission line at the red edge of the spectrum centered at $\lambda_{line}=9552$ \AA~, with signal to noise ratio of more than 10 (Fig. \ref{f:hostspec}). If interpreted as the H$\alpha$ ($\lambda 6563$ \AA) line redshifted at $z=0.455$, then
it would be difficult to explain the absence of other emission lines 
as the [OII] ($\lambda3727$ \AA) and the [OIII] ($\lambda5007$ \AA), expected to be observed at 5423 \AA~and 7285 \AA~, respectively. Another interpretation, consistent both with the lack of the Ly$\alpha$ cut-off and with the absence of any other strong emission line blue-ward of the detected one, is that we are indeed observing the [OII] ($\lambda3727$~\AA) emission line, redshifted to $z=1.563\pm0.001$. According to this interpretation, we further identify three statistically less significant features as the MgII ($\lambda2798$ \AA) and the FeII ($\lambda2586$ \AA) absorption lines and the CIII ($\lambda1909$\AA) narrow emission line.

In order to estimate the metallicity and the dominant stellar  population age, we fitted the spectrum with several template spectra. Templates are from the library of evolutionary stellar population synthesis models GALAXEV based on the isochrone synthesis code of Bruzual \& Charlot (\cite{Bruzual2003}). 
Due to the very low signal to noise ratio, we fit mainly the spectral continuum. Doing so, we are only sensitive to the age of the dominant population which fixes the slope and not to the metallicity. By fixing the metallicity to different values, we find that in any case the dominant stellar population age should be less than 0.2 Gyr.

%
%

From the [OII] line luminosity, we attempted to estimate the star formation rate (SFR) for this galaxy. We measure the strength of the line by a Gaussian fit performed with the IRAF\footnote{IRAF (Image Reduction and  Analysis Facility) is written and  supported by the IRAF programming group at  the National Optical Astronomy Observatories (NOAO)  in Tucson, Arizona. NOAO  is operated by the  Association of Universities for Research in Astronomy  (AURA), Inc. under cooperative agreement with the National Science  Foundation. IRAF is available at {\rm http://iraf.noao.edu/}.} {\it splot} package and we find $(0.57\pm0.04)\times10^{-16}$ erg cm$^{-2}$ s$^{-1}$.  Applying the relation from Kennicutt (\cite{Kennicutt1998}) we find a SFR of $(6.3\pm0.4)$ $M_{\odot} yr^{-1}$. An alternative method to measure the SFR is through the UV (1500\AA~ $\div$ 2800\AA) luminosity (Kennicutt \cite{Kennicutt1998}). The average UV flux of our spectrum is $(1.36\pm0.50)\times10^{-19}$ erg cm$^{-2}$ s$^{-1}$ \AA$^{-1}$ from which the inferred SFR is $(2.3\pm0.8)$ $M_{\odot} yr^{-1}$. We note that the latter evaluation is more robust than the former one since the UV luminosity is a direct tracer of metallicity. The derived SFR values have not been corrected for a possible local extinction and therefore $(2.3\pm0.8)$ $M_{\odot} yr^{-1}$ should be considered as the lower limit of the real SFR.

Our findings are consistent with the typical SFR range of values ($0.7\div12.7$ M$_{\odot}$ yr$^{-1}$, not corrected for extinction) found for GRBs (Christensen et al. 2004). 
Together with the age estimation obtained by $\chi^2$ minimization from the template spectra fitting, a coherent description of a  late type galaxy with moderate-high star formation activity is emerging, further strengthening the identification of this galaxy as the host of XRF 040912.

\section{Discussion}

\subsection{A ``dark'' X-ray flash}

\begin{figure}
   \centering
   \includegraphics[width=9cm]{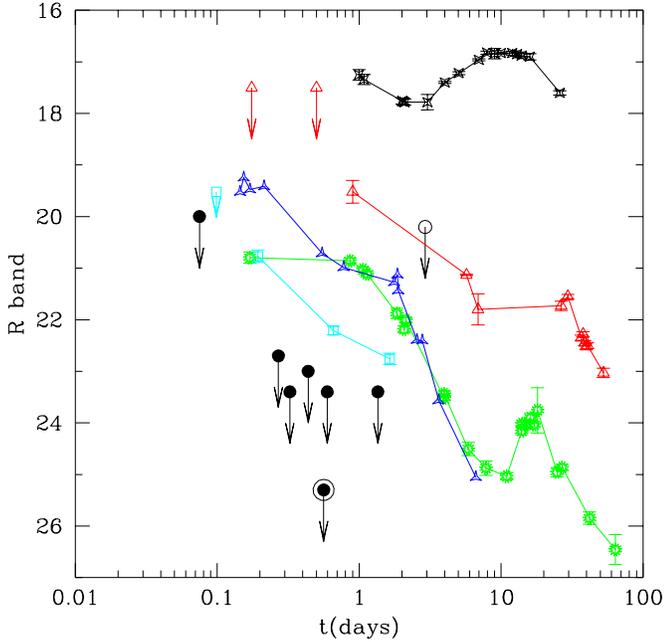}
      \caption{Optical afterglow R-band light curves for the XRFs at known redshift, up to $\sim$50 days after
the burst. The magnitudes were corrected only for Galactic extinction, but for \object{XRF~020903} (upward pointing red triangles, Bersier et al. \cite{Bersier2006}) and for \object{XRF~060218} (starred open squares) Mirabal et al. (\cite{Mirabal2006}) for which, in addition, the host galaxy subtraction was applied. The other data are from: \object{XRF~030429} (starred open triangles) Jakobsson et al. (\cite{Jakobsson2004}); \object{XRF~050416A} (open squares) Kahharov et al. (\cite{Kahharov2005}), Yanagisawa et al. (\cite{Yanagisawa2005}), Holland et al. (\cite{Holland2006}); \object{XRF~040701}, (open circle) de Ugarte Postigo et al. (\cite{deugart2004}); \object{XRF~030723} (starred open circles) Fynbo et al. (\cite{Fynbo2004}).  We put for \object{XRF~040912} (filled circles) the deepest optical upper limits for the HETE-2/WXM error-box down to which no variable source was found by the Kiso (Ogura et al. 2004), NOT (Andersen et al. 2004), TLS (this work) and Blanco (Rest et al. 2004) telescopes and the upper limit found at the position of the X-ray afterglow candidate from residual image obtained from Magellan image subtraction (marked filled circle).
}
         \label{f:Raft}
   \end{figure}

The optical flux upper limits of \object{XRF 040912} measured from different telescopes (Tab. \ref{t:ul}) are among the faintest ever obtained (Fig.\ref{f:Raft}), ranking this burst among the ``darkest" XRFs (and GRBs, see e.g. Rol et al. 2005). We note that the Galactic extinction towards this burst is small. Indeed, from Schlegel et al. (1998), the Galactic reddening is $E(B-V)=0.028$. 
The observed faintness of \object{XRF 040912} cannot be ascribed only to distance since, for example, the more distant \object{XRF~030429} shows a much brighter flux than this case. This suggests an intrinsically faint and/or rapidly fading early afterglow or a highly extinguished line-of-sight. 

The X-ray afterglow flux is among the faintest ones but still comparable to other XRF X-ray afterglows observed at similar epochs (a few days after the burst event). In particular, both the flux and decay rate are very similar to those observed for the X-ray afterglow of XRF 030723, for which however a bright optical counterpart has been detected (Fig. \ref{f:Xlc} and \ref{f:Raft}). A useful piece of information comes from the likely host galaxy of \object{XRF 040912} that shows no evidence of reddening, with a ``flat" spectral continuum (Fig. \ref{f:hostspec}). A possible interpretation of these findings is that the afterglow of this XRF is intrinsically fainter with respect to other XRFs. Another explanation which we cannot exclude is that the optical emission was suppressed by local (host galaxy) dust. The lack of reddening in the host galaxy spectrum may be explained by a localized dusty environment in the vicinity of the XRF, as suggested for \object{GRB 000210}, another 'dark' burst with similar characteristics (Gorosabel et al. \cite{Gorosabel2003}). Alternatively, the lack of reddening may be due to a weak wavelength dependence of the dust extinction law, as inferred for some GRBs and for other extra-galactic objects such as AGNs (Maiolino et al. \cite{Maiolino2000}, Stratta et al. \cite{Stratta2004b}, Chen et al. \cite{Chen2006}). In the latter case, a high equivalent hydrogen column density is expected from the X-ray data analysis, while we found a value consistent with the Galactic absorption. However, the X-ray data quality is too low to exclude a large hydrogen column density value. A simple exploration of the parameter space, leaving all the model parameters (see \S 3.3) free to vary, indicates an extra-galactic absorption upper limit of $N_H\le7.4\times10^{22}$ cm$^{-2}$ (95$\%$ confidence level) assuming a solar abundance and a cold, neutral gas. Therefore, although we cannot exclude the presence of a large amount of neutral hydrogen, we consider that this possibility is not strongly supported by the observations.

%

\subsection{Energetics}

\begin{figure}
   \centering
   \includegraphics[width=9cm]{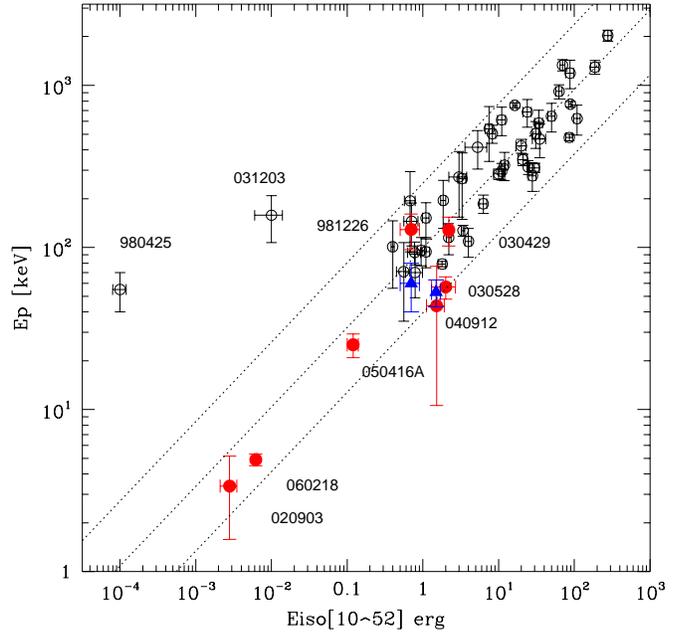}
      \caption{
The $E_{peak}-E_{iso}$ relation (Amati et al. 2002, Amati 2006). The open circles are the GRBs and XRR bursts while filled circles are the observed XRFs so far. The dotted lines are the best fit power law $E_{peak,i}=99\times E_{iso}^{0.49}$ delimitated by a logarithmic deviation of 0.4 (Amati 2006). The filled triangles are XRF 021104 and XRF 030823 for which the distance scale has been estimated through the pseudo-redshift (P\'elangeon et al. 2006).
}
         \label{f:Eiso}
   \end{figure}

The estimation of the distance scale for \object{XRF 040912} allows us to investigate the intrinsic properties of this burst. Correcting the peak energy of the $\nu F_{\nu}$ prompt emission spectrum for the cosmological redshift, we find $E_{peak,i}=44\pm33$ keV, while the total isotropic-equivalent released energy implied from the observed fluence during the prompt emission is $E_{iso}=(1.5\pm0.4)\times10^{52}$ erg. These values are consistent with the $E_{peak,i}-E_{iso}$ relation found to be valid for a large sample
of bursts (e.g. Amati et al. \cite{Amati2002}), including GRBs, XRR bursts and XRFs (Fig.\ref{f:Eiso}).

%

   \begin{figure}
   \centering
   \includegraphics[width=9cm]{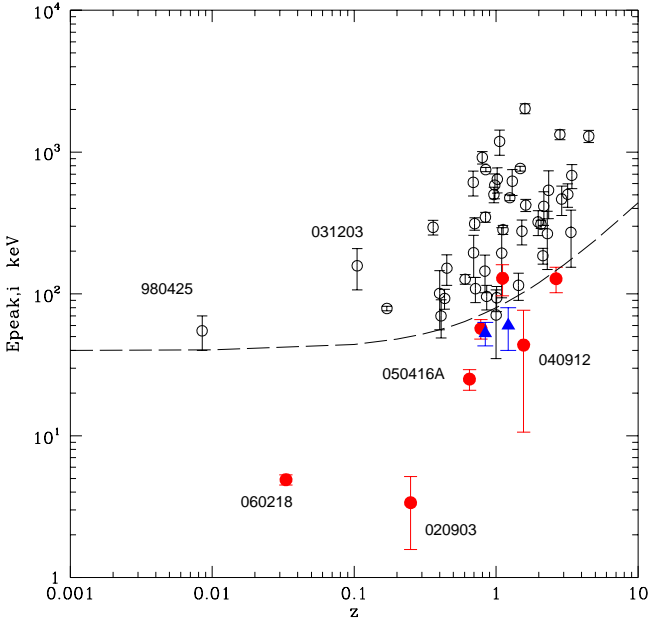}
      \caption{The intrinsic peak energy $E_{peak,i}$ of GRBs, XRR bursts (from Amati \cite{Amati2006}, open circles) and XRFs (filled circles) is plotted against redshift. The dashed line marks the intrinsic peak energy evolution with redshift, assuming a burst with {\it observed} peak energy of 40 keV.
}
         \label{Epeak_z}
   \end{figure}

However, the $E_{peak,i}$ and $E_{iso}$ values rank \object{XRF 040912} among the intrinsically soft GRBs rather than among the intrinsic XRFs, as evident in Figure \ref{f:Eiso}. A similar case was reported by Rau et al. (\cite{Rau2005}) for \object{XRF 030528} at $z=0.782$ that would have been classified as a soft GRB in its rest frame. Two more XRFs with rest frame properties more similar to soft GRBs are XRF 030429 at $z=2.65$ and \object{XRF 981226} at $z=1.11$, although for the latter, the intrinsic peak energy measure is affected by large uncertainty (D'Alessio et al. \cite{Dalessio2005}). In addition, recent studies of XRF 050215B at unknown redshift, have addressed the faintness of its prompt event to a moderate luminosity distance (Levan et al. 2006). Another intermediate case that would have been classified as an X-ray Rich burst is XRF 050416A at $z=0.65$ (Sakamoto et al. \cite{Sakamoto2005}) while the only {\it intrinsic} XRFs (hereafter {\it i}-XRFs) detected so far are XRF 020903 at $z=0.2506$ (Sakamoto et al. \cite{Sakamoto2004}) and XRF 060218 at $z=0.03342$ (Campana et al. \cite{Campana2006}), both with an intrinsic $E_{peak}$ below 10 keV.

Motivated by these findings, we further investigated the intrinsic properties of other XRFs detected by HETE-2 with no measured spectroscopic redshift but for which the pseudo-redshift was determined using a distance estimator based on the prompt emission (Atteia 2003, P\'elangeon et al. 2006). The pseudo-redshifts computed so far have provided redshift estimate within a factor of 2 for tens of bursts (P\'elangeon et al. 2006).
Its applicability is reinforced here by \object{XRF 040912}, with $z_p=2.9\pm1.6$ consistent with the spectroscopic redshift of the host galaxy. Two suitable candidates are XRF 021104 and XRF 030823 (P\'elangeon et al. 2006) for which the pseudo-redshifts are $z_p=1.2\pm1.1$ and $z_p=0.8\pm0.7$, respectively. For XRF 021104 we find $E_{iso} = (0.7\pm0.2) \times 10^{52}$ erg  and $E_{peak} =  60^{+27}_{-15}$ keV while for XRF 030823 we find $E_{iso} = (1.5\pm0.2) \times 10^{52}$ erg  and $E_{peak} =  53^{+11}_{-9}$ keV. These values again classify these two bursts among the intrinsic soft GRBs rather than among the {\it i}-XRFs (Fig. \ref{f:Eiso}, filled triangles). We stress that the pseudo-redshift is an 'estimator' of the distance scale and this result cannot be certain, although it is still a plausible explanation.

\section{The nature of X-Ray Flashes}

These findings point out that the first interpretation of XRFs as high-redshift GRBs is turning out to be correct if XRFs are interpreted as {\it soft GRBs at intermediate redshifts}. This is true for all XRFs detected so far, but two. For the two {\it i}-XRFs 060218 and 020903 another explanation is required. It is worth noting, however, that both GRBs and {\it i}-XRFs nicely satisfy the $E_{peak,i}-E_{iso}$ relation, indicating a possible common origin of the prompt emission mechanism.

We further attempt to extract useful information by comparing the intrinsic properties of XRFs, XRR bursts and GRBs. We compare \object{XRF 040912} with 45 long bursts taken from the burst sample quoted by Amati (\cite{Amati2006}), for which good estimates of redshift and observed peak energy were obtained.
In particular, we consider the $E_{peak,i}$ value of GRB 031203 as quoted by Amati (\cite{Amati2006}), although this
value is still uncertain (see also Watson et al. \cite{Watson2006}). In Figure \ref{Epeak_z} we plot the intrinsic peak energy as a function of redshift. In this plot, the $E_{peak,i}$ values of XRFs are the natural extension of the $E_{peak,i}$ of GRBs  towards soft values, forming a single population of bursts with 20 keV $<E_{peak,i}<$ 2000 keV, where those bursts with {\it observed} peak energy below 40 keV are identified as XRFs. Two possible outliers are the two nearby {\it i}-XRFs (namely XRF 020903 and XRF 060218). Although at this stage we can not exclude a continuum distribution down to very soft bursts, the gap between the majority of the bursts and these two nearby {\it i}-XRFs may possibly indicate a different population of very soft bursts (see also Liang et al. \cite{Liang2006}). 
A distinct population of soft and sub-energetic bursts has been recently proposed by Mazzali et al. (\cite{Mazzali2006}) where the diversity from the brighter and harder bursts stems from a lower mass progenitor. This scenario was driven not only by the softness and the faintness of the prompt emission of XRF 020903 and XRF 060218, but also by the peculiar  properties of their associated SNe with respect to other GRB-SNe. Despite low mass progenitors (20-25 $M_{\odot}$) possibly being more common than high mass (35-50 $M_{\odot}$) ones, the softness of their bursts would make their detection more difficult than for the harder bursts due to the spectral coverage presently afforded by the flying high energy satellites. Similar conclusions have been reached by Liang et al. (\cite{Liang2006}) from GRB luminosity function studies.
The {\it i}-XRFs segregation is however less evident in the $E_{iso}$ (or $L_{iso}$) versus redshift diagram (see e.g. Liang et al. 2006).
Further identifications of {\it i}-XRFs will help answer the question of whether the soft peak energy bursts form a distinct population or if they are rather the extension to soft energies of a single phenomenon.

\section{Summary and conclusions}

We performed a multi-wavelength analysis of XRF 040912. We find no optical afterglow counterpart. The limiting magnitude, computed from image subtraction, ranks this burst among the 'darkest' ever detected so far, with $R>25.3$ mag at 13.6 hours after the burst. Among 22 X-ray sources detected by Chandra in the HETE-2 error-box, only one shows a statistically significant fading behavior with decay rate consistent with the typical values measured for X-ray afterglows at similar epochs (few days after the burst). No evidence of any SN re-brightening has been detected 22.3 days after the burst. The non-fading optical source positionally consistent with the X-ray afterglow candidate, likely the host galaxy, shows a spectrum with a single emission line at 9550~\AA. The lack of any other strong emission lines blue-ward of the detected one and the absence of the Ly$\alpha$ cut-off down to 3800~\AA~ are consistent with the hypothesis of the [OII] line at redshift $z=1.563\pm0.001$. The intrinsic spectral properties rank this XRF among the soft GRBs in the $E_{peak,i}-E_{iso}$ diagram. Similar results are obtained for most XRFs at known redshift (or pseudo-redshift), suggesting the existence of two types of XRFs: one type is made by soft GRBs at high ($z\sim1$) redshifts (most XRFs are of this type) and another type ({\it i}-XRFs), as XRF 060218 and XRF 020903, that show intrinsic soft properties possibly associated with a different progenitor population. These observations,  may call for a new definition of XRFs, which would be restricted to transients with an $E_{peak}$ of a few keV. Such a definition is however beyond the scope of this paper.

\begin{acknowledgements}
The authors are grateful to Thierry Contini and Jacob Walcher for useful discussion and to the anonymous referee for  helpful suggestions.
G.S. and B.G. acknowledge the support by the EU Research and Training Network "Gamma-Ray Bursts, an Enigma and a Tool."
This work is based in part on observations obtained with MegaPrime/MegaCam, a joint project of CFHT and CEA/DAPNIA, at the Canada-France-Hawaii Telescope (CFHT) which is operated by the National Research Council (NRC) of Canada, the Institut National des Science de l'Univers of the Centre National de la Recherche Scientifique (CNRS) of France, and the University of Hawaii and on  observations collected  at  the European Southern Observatory (ESO),  Chile,  under Director's Discretionary Time Proposal 275.A-5041.

\end{acknowledgements}

\end{document}